%% file: main.tex
\documentclass[sigconf, nonacm]{acmart}
\usepackage{xspace} 
\usepackage{booktabs}
\usepackage[noend]{algpseudocode}
\usepackage{enumitem}
\usepackage{graphicx}

\newcommand\vldbdoi{XX.XX/XXX.XX}
\newcommand\vldbpages{XXX-XXX}
\newcommand\vldbvolume{14}
\newcommand\vldbissue{1}
\newcommand\vldbyear{2020}
\newcommand\vldbauthors{\authors}
\newcommand\vldbtitle{\shorttitle} 
\newcommand\vldbavailabilityurl{http://vldb.org/pvldb/format_vol14.html}
\newcommand\vldbpagestyle{plain}

\usepackage{ifthen}
\newboolean{publicversion}
\setboolean{publicversion}{false}
\ifthenelse{\boolean{publicversion}}{
  \newcommand{\grumbler}[3]{}
}{
  \newcommand{\grumbler}[3]{\textcolor{#3}{\bf #1: #2}}
}

\input{macros}

\hypersetup{
  colorlinks=true,
  linkcolor=blue,
  filecolor=blue,
  urlcolor=blue,
  citecolor=blue,
}

\begin{document}
\title{Rainblock: Faster Transaction Processing in Public Blockchains}

\author{Soujanya Ponnapalli}
\affiliation{%
  \institution{University of Texas at Austin}
}

\author{Aashaka Shah}
\affiliation{%
  \institution{University of Texas at Austin}
}

\author{Amy Tai}
\affiliation{%
  \institution{VMware Research}
}

\author{Souvik Banerjee}
\affiliation{%
  \institution{University of Texas at Austin}
}

\author{Dahlia Malkhi}
\affiliation{%
  \institution{Novi Financial}
}

\author{Vijay Chidambaram}
\affiliation{%
  \institution{University of Texas at Austin}
  \institution{VMware Research}
}

\author{Michael Wei}
\affiliation{%
  \institution{VMware Research}
}

\input{abstract2}

\maketitle

\pagestyle{\vldbpagestyle}
\begingroup\small\noindent\raggedright\textbf{PVLDB Reference Format:}\\
\vldbauthors. \vldbtitle. PVLDB, \vldbvolume(\vldbissue): \vldbpages, \vldbyear.\\
\href{https://doi.org/\vldbdoi}{doi:\vldbdoi}
\endgroup
\begingroup
\renewcommand\thefootnote{}\footnote{\noindent
This work is licensed under the Creative Commons BY-NC-ND 4.0 International License. Visit \url{https://creativecommons.org/licenses/by-nc-nd/4.0/} to view a copy of this license. For any use beyond those covered by this license, obtain permission by emailing \href{mailto:info@vldb.org}{info@vldb.org}. Copyright is held by the owner/author(s). Publication rights licensed to the VLDB Endowment. \\
\raggedright Proceedings of the VLDB Endowment, Vol. \vldbvolume, No. \vldbissue\ %
ISSN 2150-8097. \\
\href{https://doi.org/\vldbdoi}{doi:\vldbdoi} \\
}\addtocounter{footnote}{-1}\endgroup

\ifdefempty{\vldbavailabilityurl}{}{
\vspace{.3cm}
\begingroup\small\noindent\raggedright\textbf{PVLDB Artifact Availability:}\\
The source code, data, and/or other artifacts have been made available at \url{\vldbavailabilityurl}.
\endgroup
}

\input{intro15}
\input{background}
\input{motivation}
\input{design2}
\input{dsm3}
\input{discussion}
\input{impl}
\input{eval}
\input{related}
\input{conclusion}

\bibliographystyle{ACM-Reference-Format}
\bibliography{main}

\end{document}

%% file: macros.tex
\newcommand{\mycaption}[2]{\caption{\textbf{#1}. {#2}}}
\newcommand{\sref}[1]{\S\ref{#1}}
\newcommand{\vheading}[1]{\vspace{0.05in}\noindent\textbf{#1}}

\newcommand{\ie}{\textit{i.e.,}\xspace}
\newcommand{\eg}{\textit{e.g.,}\xspace}

\newcommand{\myx}{$\times$\xspace}
\newcommand{\vtt}[1]{\texttt{#1}\xspace}
\newcommand{\sysname}{{\small\textsc{RainBlock}}\xspace}

\newcommand{\dsm}{{\small\textsc{DSM-Tree}}\xspace}
\newcommand{\dsms}{{\small\textsc{DSM-Trees}}\xspace}
\newcommand{\dsmfull}{{Distributed, Sharded Merkle Tree}\xspace}

\newcommand{\dsmtree}{DSM-Tree\xspace}
\newcommand{\Dsmtree}{DSM-Tree\xspace}

\newcommand{\io}{I/O\xspace}

%% file: abstract2.tex
\begin{abstract}
  Public blockchains like Ethereum use Merkle trees to verify
  transactions received from untrusted servers before applying them to
  the blockchain. We empirically show that the low throughput of such
  blockchains is due to the I/O bottleneck associated with using
  Merkle trees for processing transactions. We present \sysname, a new
  architecture for public blockchains that increases throughput
  without affecting security. \sysname achieves this by tackling the
  I/O bottleneck on two fronts: first, decoupling transaction
  processing from I/O and removing I/O from the critical path; second,
  reducing I/O amplification by customizing storage for
  blockchains. \sysname uses a novel variant of the Merkle tree, the
  \dsmfull (\dsm) to store system state.  We evaluate \sysname using
  workloads based on public Ethereum traces (including smart
  contracts) and show that \sysname processes 20K transactions per
  second in a geo-distributed setting with four regions spread across
  three continents.
\end{abstract}

%% file: intro15.tex
\section{Introduction}
\label{sec-intro}

Blockchains are decentralized systems that maintain an immutable history
of transactions as a chain of blocks; each block has an ordered list of transactions. Public blockchains allow untrusted servers to process transactions, while private or permissioned blockchains only allow a few
specific servers. The decentralized nature,
transparency, fault tolerance, and auditability of public blockchains
have led to many applications in a range of domains like
crypto-currencies~\cite{bitcoin, ethereum, nxt-crypt},
games~\cite{evans2019cryptokitties}, and
healthcare~\cite{mcghin2019blockchain}.

\vheading{The problem: low throughput}.  Public blockchains suffer
from low throughput. The two widely-used public blockchains,
Bitcoin~\cite{bitcoin} and Ethereum~\cite{ethereum}, can process only
tens of transactions per second, severely limiting their
applications~\cite{wongcryptokitties}. The low throughput is due to
the design choices that ensure
security~\cite{Jakobsson1999ProofsOW}. A new block can only be created
after a majority of servers receive and process the previous
block~\cite{12s-eth}. However, simply adding more transactions per
block will not increase throughput as it increases the block
processing time and reduces the block creation rate. One way to safely
increase throughput is to first increase the transaction processing
rate at servers and then pack more transactions per block, without
affecting the block creation rate. We analyze transaction processing
at servers and find that it has \io bottlenecks.

\vheading{Case study: Ethereum}. We analyze block processing time in
the widely-used Ethereum blockchain. Servers in Ethereum that add
blocks to the blockchain are termed miners. Ethereum miners store the
system state using a Merkle tree \cite{merkle1987digital,
  ethereum-mpt} on local storage. As a result, processing transactions
requires performing disk \io to read and update multiple values in the
Merkle tree. We highlight two main problems with how transactions are
processed in Ethereum:
\begin{itemize}[leftmargin=*]
\item \vheading{\io Amplification}. The Merkle tree is stored using
  the RocksDB key-value store~\cite{rocksdb}, and Merkle tree nodes
  are looked-up using their cryptographic hashes. As hashes randomize
  node locations on storage, traversing the tree requires performing
  random reads~\cite{raju18mlsm}. The log-structured merge
  tree~\cite{o1996log} at the heart of RocksDB further causes \io
  amplification~\cite{RajuEtAl17-PebblesDB}. Processing a single block
  of 100 transactions in Ethereum requires performing more than 10K
  random \io operations (100\myx higher) and takes hundreds of
  milliseconds even on a datacenter-grade NVMe SSD.
\item \vheading{\io in the critical path}. Servers process
  transactions one by one, performing slow \io operations on local
  storage in the critical path. This design limits the \io parallelism
  in the system and lowers the overall transaction processing rate.
\end{itemize}  

This \io bottleneck in Ethereum prevents servers from packing more
transactions per block~\cite{yang2019empirically} and results in the
low transaction throughput. Researchers have noted similar \io
bottlenecks in other public blockchains \cite{yang2019prism,
  reyzin2017improving}.

\vheading{Prior Work}.  While researchers have attempted to increase
throughput by designing new consensus protocols~\cite{Gilad:algorand,
  bitcoin-ng, Miller:2016:honeybadger, 203704, ethereum-casper}, the
problem of processing transactions without \io bottlenecks is
orthogonal to how consensus is achieved, and remains unsolved in these
protocols. Another line of work shards the blockchain to reduce \io
inside each
shard~\cite{omniledger,Zamani:rapidchain,Luu:Elastico,227661}, but
sharding weakens the security guarantees of public
blockchains~\cite{sonnino2019replay}. Finally, researchers have
proposed new data structures to reduce the \io
bottleneck~\cite{reyzin2017improving}; while this reduces the amount
of \io performed, \io is still performed in the critical path, leading
to low throughput.

\vheading{Main Idea: faster transaction processing}. This work aims at
increasing the throughput of public blockchains by removing \io
bottlenecks and increasing transaction processing rate. It tackles the
\io bottleneck on two fronts: first, by avoiding \io in the critical
path, and second, reducing \io amplification by building a customized
Merkle tree-based storage for blockchains.

\input{fig-compact-arch}

\vheading{Our system}.  This paper presents \sysname, a new
architecture for building public blockchains, that increases
throughput by making transaction processing faster. \sysname does not
change the consensus protocol or the block creation rate, and hence is
able to increase throughput without weakening security
guarantees. \sysname achieves high throughput by removing the \io
bottlenecks in transaction processing. \sysname introduces two novel
aspects: an architecture that decouples \io from transaction processing
and removes \io from the critical path, and a variant of the
Merkle tree that reduces \io amplification.  

\vheading{The \sysname architecture}. \sysname decouples \io from
transaction processing. \sysname achieves this by introducing
\emph{clients} that prefetch data and Merkle proofs (that show that
the data is correct) from \emph{storage nodes} and submit them to the
miners. The miners do not perform \io operations, and process
transactions using the information provided by
clients. Figure~\ref{fig-compact-arch} illustrates the
architecture. The key is that \io is moved outside the critical path,
since miners can now process transactions without performing additional
\io. The architecture also increases \io parallelism, since multiple
clients can perform \io in parallel, and multiple miners can update
storage in parallel. \sysname stores state in the \dsm data structure.

\vheading{The \dsmfull (\dsm)}. \sysname reduces \io amplification
by using a novel variant of the Merkle tree. The \dsm is an in-memory,
sharded, multi-versioned, two-layered Merkle tree. It reduces \io
amplification by using an efficient in-memory representation of the
Merkle tree. Traversing the tree does not require expensive hashing
operations. The \dsm is sharded as the entire Merkle tree will not fit
in the DRAM of a commodity server; requiring servers with high DRAM
capacity would reduce the decentralization of \sysname. The \dsm has a
bottom layer and a top layer. The bottom layer consists of the Merkle
tree sharded across storage nodes. The bottom layer is
multi-versioned: when a miner updates storage, a new version is
created in a copy-on-write manner. Each miner has a private top layer
that represents a single version or snapshot of the system state. The
miner executes transactions against the snapshot in the top layer. The
top layer provides consistent reads, regardless of concurrent updates
and multiple versions at the bottom layer.

\vheading{Challenges}. The \sysname architecture has to solve several
challenges to be effective. First is consistency in the face of
concurrent updates. This is solved by multi-versioning in \dsm.  Since
concurrent updates from miners create new versions (rather than
modifying existing data), clients always read consistent data. A
second challenge is prefetching data for Turing-complete smart
contracts.  With smart contracts that can execute arbitrary code, it
is not straightforward for the clients to prefetch all the required
data. \sysname solves this by having clients \emph{speculatively
  pre-execute} the transaction to obtain data and proofs. A third
related challenge is that clients may submit stale proofs to the
miners (miners may update storage after clients finish
reading). \sysname handles this by having miners tolerate stale proofs
whenever possible, via \emph{witness revision}, allowing transactions
that would otherwise abort to execute. Finally, \sysname trades local
storage \io for network \io: the network may become the new
bottleneck. \sysname handles this by deduplicating data and proofs
before sending them over the network, and exploiting the synergy
between the top and bottom layers of the \dsm to reduce proof sizes as
much as possible.

\vheading{Implementation and Evaluation}. We implement \sysname
prototype by modifying Ethereum. We chose Ethereum as the
implementation base and comparison point for two main reasons. First,
Ethereum has been in operation as a public blockchain for nearly six
years, presenting a large amount of data to test our
assumptions. Second, Ethereum supports Turing-complete smart
contracts~\cite{szabo1994smart}, allowing the codification of complex
decentralized applications.

To evaluate \sysname, we generate synthetic workloads that mirror
transactions on Ethereum mainnet. We analyzed Ethereum transactions
and observed that user accounts involved in transactions have a Zipf
distribution: 90\% of transactions involve the same 10\% of user
accounts. We observed that only 10--15\% of Ethereum transactions
involved smart contracts. Our workload generator faithfully reproduces
these distributions. We evaluated \sysname using these workloads and
found that Rainblock processes 20\myx more transactions than Ethereum
in a geo-distributed setting with four regions spread across three
continents. \sysname processes 30K transactions per second in a single
node setting, and 20K transactions per second in the geo-distributed
scenario.

In summary, this paper makes the following contributions:
\begin{itemize}
\item To the best of our knowledge, the first system
  architecture that eliminates the \io bottleneck in public
  blockchains, while supporting smart contracts.
  \item The novel \dsm authenticated data structure.
  \item  A workload generator for Ethereum transactions
    based on an analysis of Ethereum transactions.
  \item Empirical evidence that \sysname is able to process 30K
    transactions per second in a single node setting, and 20K
    transactions per second in a geo-distributed setting.
\end{itemize}  

%% file: fig-compact-arch.tex
\begin{figure}
  \includegraphics[width=\columnwidth]{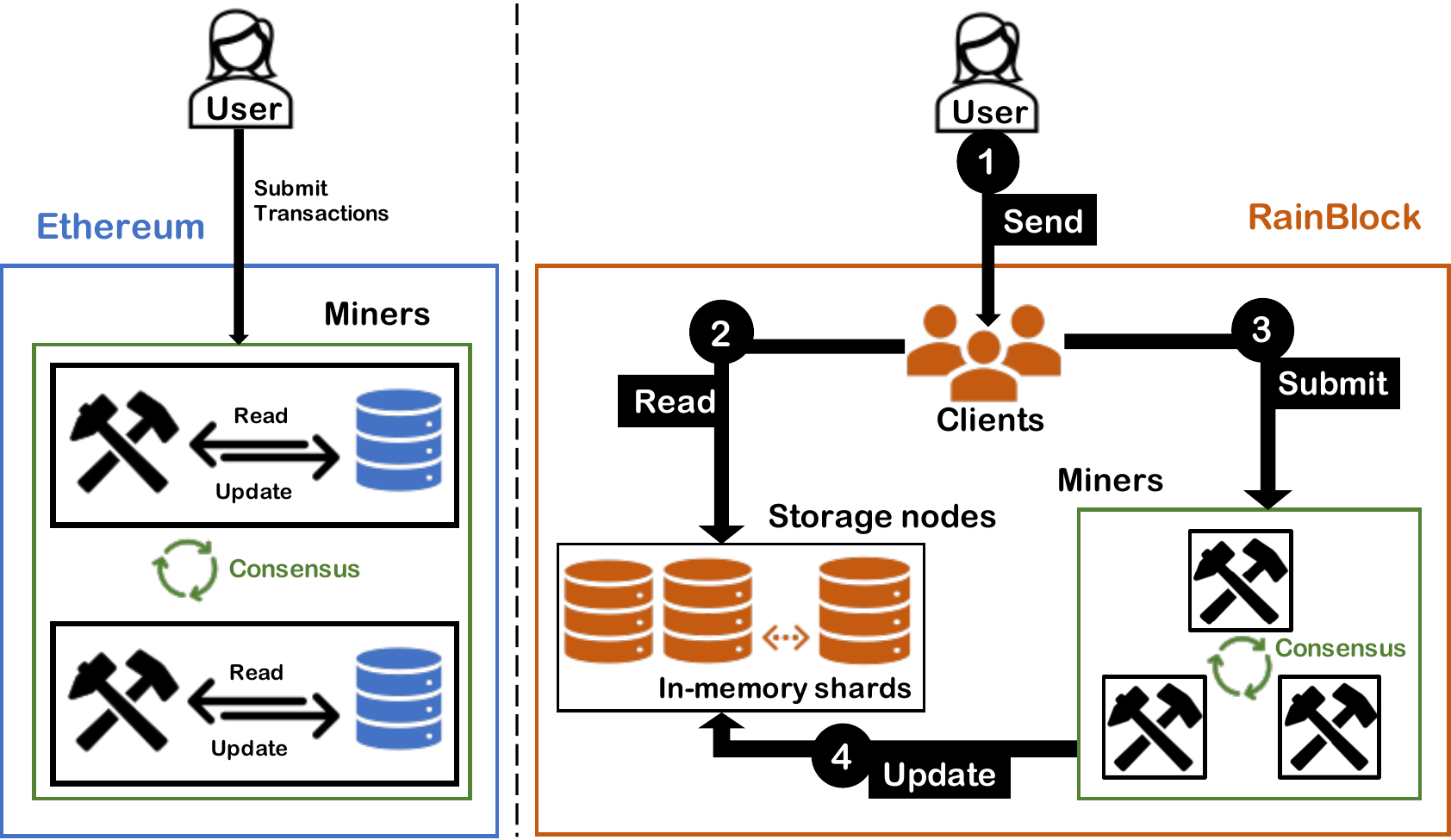}
  \vspace{-10pt}
  \mycaption{Ethereum and \sysname architecture}{Miners in Ethereum perform local disk \io in the critical path. In \sysname, clients read data from remote in-memory storage nodes (out of the critical path) on behalf of its miners. Miners execute txns without extra \io and update storage nodes.
  }
  \label{fig-compact-arch}
\end{figure}

%% file: background.tex
\section{Background}
\label{sec-background}

In this section, we provide some background on public blockchains and
discuss how public blockchains use Merkle trees.

\subsection{Public Blockchains}
\label{sec-public-blockchains}

Public blockchains, like databases, maintain some system state and
support transactions on that system state. Public blockchains are
widely used because of their open networks, decentralized
architectures, and immutable, auditable state.

\vheading{Open Networks}.  Public blockchains allow untrusted servers
to join their network and process transactions. These untrusted
servers are responsible for storing and advancing the system state and
the blockchain. Every block in the blockchain has an ordered list of
transactions. Further, blocks store cryptographic hashes of their
previous blocks, creating a chain of blocks that is cryptographically
secure and immutable. In public blockchains, the system state advances
from one snapshot to the other with every new block of transactions,
providing consistent snapshots of the system state.

\vheading{Decentralization}.  In public blockchains, the untrusted
servers in their networks can be malicious and can provide incorrect
information about the latest system state. Thus, public blockchains
follow the State Machine Replication (SMR)~\cite{smr, paxos} paradigm
to tolerate a minority of malicious servers. Each untrusted server in
the network acts as a state machine replica that starts from a fixed
initial state. Following SMR, every non-malicious server that begins
with the same initial state and processes the same blocks of
transactions, arrives at the same final system state. Public
blockchains rely on this non-malicious quorum to serve the correct
system state after processing the transactions in the blockchain.

\vheading{Consensus}.  In public blockchains, servers that create new
blocks and extend the blockchain (termed miners) can also be
malicious. As a result, public blockchains rely on consensus protocols
like Proof-of-Work (PoW)~\cite{Jakobsson1999ProofsOW} to maintain
their correctness and liveness. Broadly, PoW ensures that miners
create a new block (roughly once every 12 seconds) only after a majority
(about 95\%) of
the servers in the network receive and process the previous
block~\cite{12s-eth}. This rate limit avoids forking of the
blockchain, \ie it restrains multiple miners from creating new blocks
on top of different previous blocks. Thus, PoW prevents forks that
ambiguate the latest system state or halt its progress, maintaining
the security and liveness of public blockchains.

\vheading{The quest for higher throughput}.
As PoW limits the block creation rate, it significantly reduces the throughput of blockchains. PoW typically allows servers to receive newer blocks after they process the previous block. As the system state increases over time, servers take more time to process a new block of transactions, further slowing down public blockchains. There has been active research on increase the throughput of public blockchains by sharding the blockchain~\cite{omniledger,Zamani:rapidchain,Luu:Elastico,227661} or with alternatives for PoW~\cite{Gilad:algorand, bitcoin-ng, Miller:2016:honeybadger, 203704, ethereum-casper}. Our work \sysname increases the throughput of public blockchains by reducing the time taken to process a single transaction and enabling larger blocks without changing the block creation rate. \sysname proposes a new architecture for public blockchains and relies on the \dsm data structure to efficiently handle the increasing system state over time.

\input{fig-mpt}
\input{fig-comb-overheads.tex}

\subsection{Merkle Trees}
\label{sec-merkle-trees}

In public blockchains, with untrusted servers maintaining the system state,
users cannot trust the data they receive from such servers. Many public
blockchains~\cite{ethereum,hyperledger-sawtooth,nxt-crypt} use
\emph{authenticated data structures} such as Merkle
trees~\cite{merkle1987digital} to prove that this data is correct.


\vheading{State authentication}. With authenticated data structures,
users can read data along with proofs (called witnesses) from
untrusted servers. These witnesses allow users to verify if the data is
correct and if it belongs to the latest snapshot of the system state.
Similarly, new servers can request the system state from an
untrusted server and verify its correctness, without having to
reconstruct it locally by replaying the entire blockchain.

\vheading{Merkle tree}.  To authenticate the system state, a public
blockchain like Ethereum~\cite{ethereum}
relies on a variant of the Merkle tree called the Merkle Patricia trie
(MPT)~\cite{ethereum-mpt}. In Ethereum, the system state is a
key-value mapping from unique addresses to user accounts. The MPT
stores these accounts in the leaf nodes and indexes them with their
addresses, as shown in Figure-\ref{fig:MerklePatriciaTree}. In this paper,
we use the terms MPT and Merkle tree interchangeably.

\vheading{Merkle root}.  In a Merkle tree, every non-leaf node stores
the cryptographic hashes of all its children. Thus, the hash of the
root node, called the Merkle root, hashes of all the values in the
system state, effectively summarizing a snapshot of the system
state. In Ethereum, as the system state changes with every
new block of transactions, each block stores a Merkle root. The Merkle
root in each block represents the expected system state after
executing all the transactions in that block.

\vheading{Merkle proof or witness}.
Witnesses allow users to identify stale or incorrect data from untrusted
servers. A witness is a vertical path in the Merkle tree
and has all the nodes from the root to the leaf storing the data.
To verify the data, users recompute a Merkle root locally using its
witness and cross-check if that Merkle root matches with the Merkle root
published in the latest block. A Merkle root mismatch indicates that the
data is stale or incorrect.

%% file: fig-mpt.tex
\begin{figure}
  \includegraphics[width=\columnwidth]{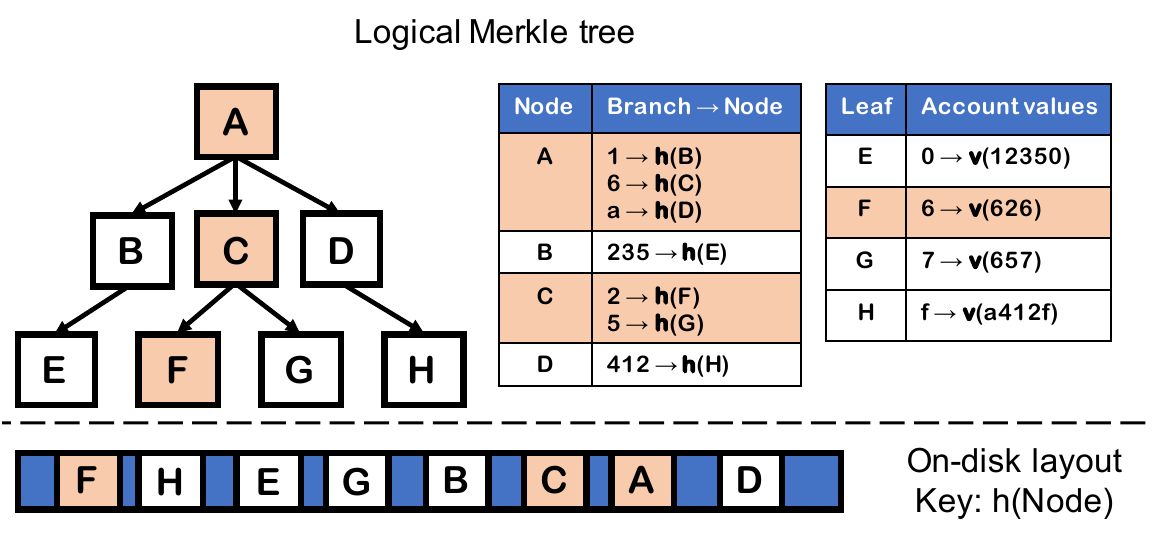}
  \mycaption{Merkle Patricia Trie}{Reading the account of an address
  $626$, for example, would require reading node $A$, traversing branch $6$, and looking up $C$ using $h(C)$. Then, traversing branch $2$ and looking up F using $h(F)$. Notice that the nodes in the MPT at random locations on disk.}
  \vspace{-10pt}
    \label{fig:MerklePatriciaTree}
\end{figure}

%% file: fig-comb-overheads.tex
\begin{figure*}[t]
    \centering
    \begin{minipage}[b]{.3\textwidth}\ContinuedFloat
        \captionsetup{labelformat=empty}
        \includegraphics[width=\columnwidth]{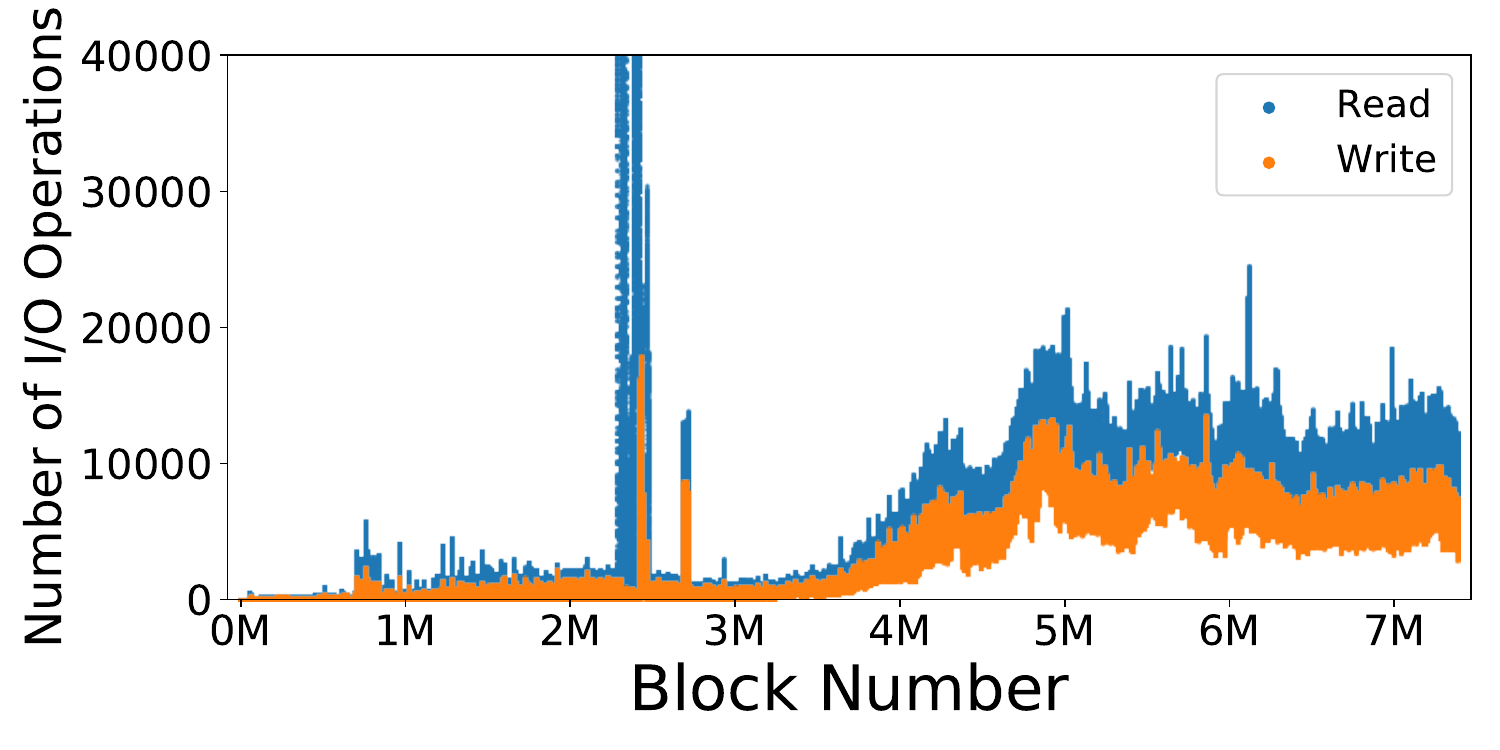}
        \caption{\emph{(a) Number of \io operations}}
    \end{minipage}\qquad
    \begin{minipage}[b]{.3\textwidth}\ContinuedFloat
        \captionsetup{labelformat=empty}
        \includegraphics[width=\columnwidth]{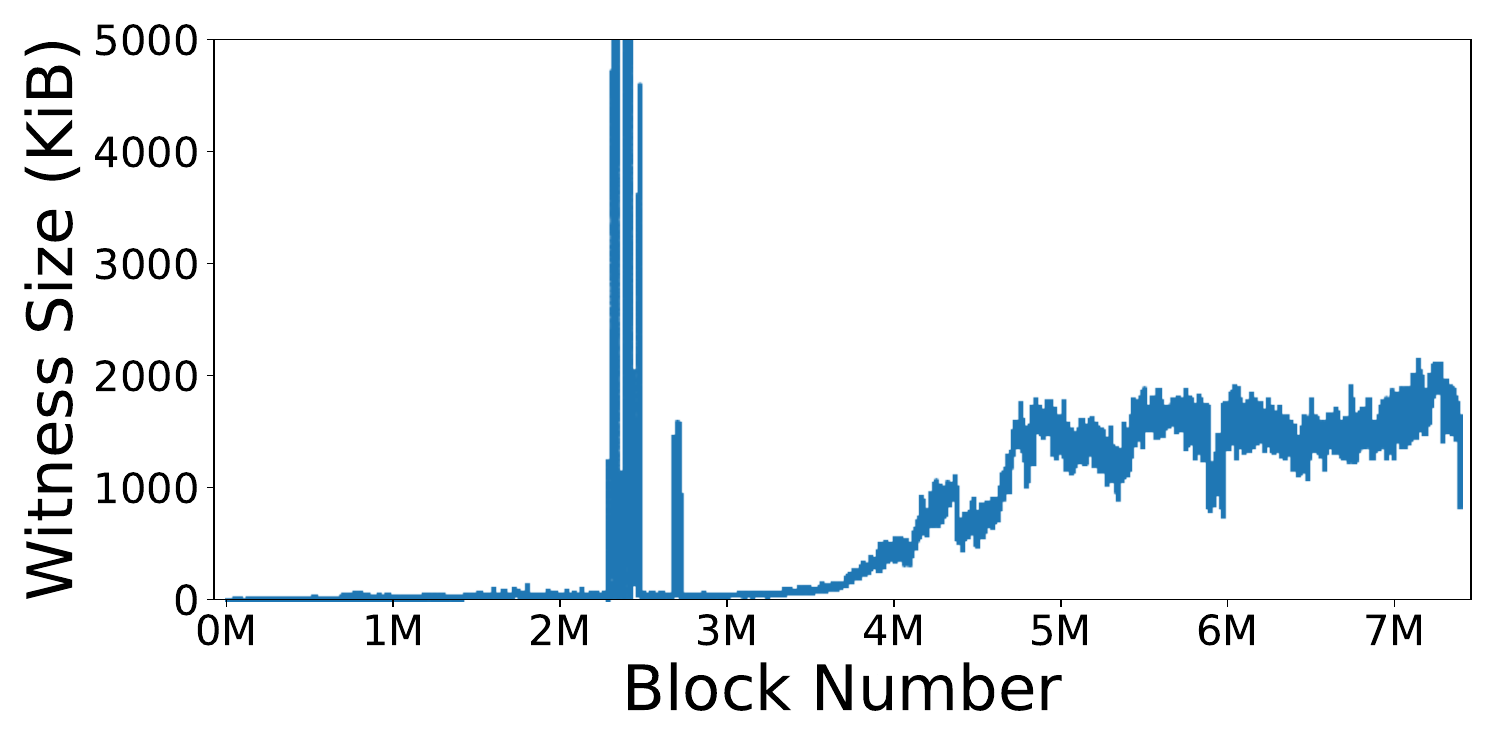}
        \caption{\emph{(b) Witness sizes}}
    \end{minipage}\qquad
    \begin{minipage}[b]{.3\textwidth}\ContinuedFloat
        \captionsetup{labelformat=empty}
        \includegraphics[width=\columnwidth]{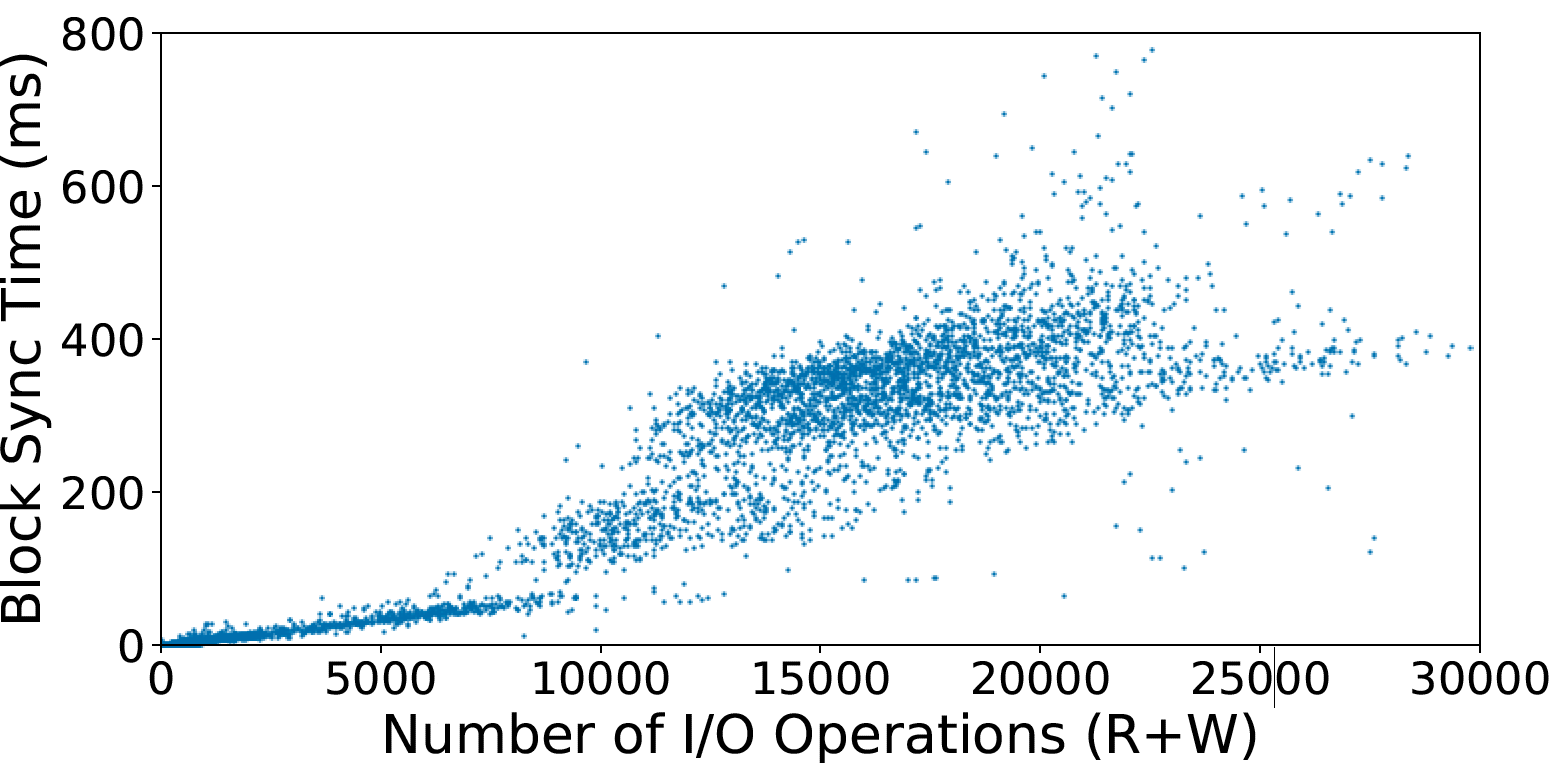}
        \caption{\emph{(c) Block processing latency}}
    \end{minipage}
    \caption{
        Overheads of the MPT. This figure highlights the \io bottleneck
        due to Merkle trees. (a) First, it shows the number of IO
        operations performed per block. (b) Then, measures the size of witnesses
        (represents the amount of data read) per block. (c) Finally, it shows the
        increase in block processing time with the increasing number of \io operations
        (\io bottleneck).}
    \label{fig-comb-overheads}
\end{figure*}

%% file: motivation.tex
\section{Motivation}
\label{sec-motivation}

We empirically show that public blockchains that use Merkle trees
to store their system state suffer from \io bottlenecks in transaction processing.
We also discuss a few strawman solutions and their drawbacks for motivating our work.

\subsection{I/O Bottlenecks}
\label{sec-io-bottleneck}

Processing a transaction involves reading and updating multiple
values in the system state. In public blockchains, once a server
receives a new block, it processes the transactions and also verifies
if its Merkle root matches the Merkle root in that block. We study the
\io bottlenecks in transaction processing using Ethereum.

The \io bottleneck in Ethereum arises from two sources. First, reading
or writing nodes of the Merkle tree generates many random I/Os in a
pointer-chasing fashion (that prevents pre-fetching). Second, the
Ethereum Merkle tree is stored on-disk using the RocksDB
key-value~\cite{rocksdb} that has inherent \io amplification
\cite{RajuEtAl17-PebblesDB,raju18mlsm}.

\vheading{Empirical study}.  We use Parity 2.2.11~\cite{parity2211}, a
popular Ethereum blockchain client, for studying the \io bottleneck
and \io amplification from Merkle trees. We use Parity to initialize a
new server that joins the Ethereum network, and replays the blockchain
(until 7.3 Million blocks) to measure various costs. We analyze the
performance impact of the Merkle tree in terms of the number of \io
operations performed and witness sizes. We also study the data
structure's effect on the block processing time.

\vheading{Number of \io operations}.  For processing a single block
with around 100 transactions, Ethereum requires performing more than
10K random \io operations (two orders of difference). Most of these
\io operations are performed for reading and updating the Merkle
tree. Figure-\ref{fig-comb-overheads}(a) shows the number of \io
operations incurred per block while processing the first 7.3 Million
block of transactions. This result shows the combined overheads from
the Merkle tree and its on-disk layout using RocksDB.

\vheading{Witness sizes}.  Witness sizes represent the amount of data
read and modified per block while reading and updating paths in the
Merkle tree. In Ethereum, which uses secure 256-bit cryptographic
hashes, the witness size of a single 100 byte user account (or value)
can be above 4 KB, showing a 40-60\myx overhead. The witness size also
increases as the total data in the Ethereum state increases, as shown
in Figure-\ref{fig-comb-overheads}(b).

\vheading{Block processing time}.  We measure the time taken to
process each block, \ie executing the transactions in that block, and
verifying if the resultant local Merkle root matches with the Merkle
root in that block. Thus, the block processing time is affected by the
\io overheads from Merkle trees. For example, processing an Ethereum
block with about 100 transactions takes hundreds of milliseconds even
on a datacenter-grade NVMe SSD. Overall,
Figure~\ref{fig-comb-overheads}~(c) shows the direct correlation
between the time taken to sync or process Ethereum blocks against the
number of IO operations required, indicating the effect of \io
bottlenecks on block processing time.

\vheading{Summary}.  Merkle trees and their on-disk key-value layout
introduces significant \io overheads that directly impact the block
processing time. Notice that our results are optimistic estimates, as
we use a datacenter-grade NVMe SSD which is probably much better hardware than that
available at an average untrusted server in the network. In
Figure-\ref{fig-comb-overheads} (a) and (b), the spikes are the result
of a DDOS attack~\cite{state-ddos-attack} on the Ethereum's state,
which creates dummy user accounts to increase the values in the Merkle
tree and thereby increases the number of \io operations and witness
sizes. Finally, these overheads will increase as the system state
increases and, as of April 1, 2019~\cite{full-state-size}, the
Ethereum state is already above 200 GB. Although we analyze Ethereum
and MPT in this study, our analysis is generally applicable to other
authenticated data structures and blockchain systems. The \io bottlenecks
from authenticated dynamic dictionaries~\cite{reyzin2017improving}
are also seen in multi-token blockchain systems such as the
Nxt cryptocurrency [nxt]~\cite{nxt-crypt}.


\subsection{Strawman Solutions}
\label{sec-strawman-solutions}

\vheading{Storing state in memory}. Can every server store the entire
state in memory to eliminate the \io bottlenecks? This would not work
as public blockchains seek to allow commodity servers to join their
networks for increasing decentralization. If servers need to
have 100s of GB of DRAM to join the public blockchain network, it
decreases decentralization. Such a solution cannot be adopted.

\vheading{Increase block size}. Can we keep the current block creation
rate and simply increase the number of transactions in each block?
This would not work since block creation rate in PoW blockchains
depends upon block processing times; with more transactions, block
processing time would increase, resulting in lower block creation rate
or causing forks that compromise security and liveness.

\vheading{Alternative consensus protocols}. Tackling the \io bottlenecks
in transaction processing is orthogonal to the underlying consensus protocol.
Even with alternative consensus protocols that release new blocks of
transactions at a much higher rate, as transaction processing happens
in the critical path, \io bottlenecks will continue to limit the
throughput~\cite{yang2019prism}.

\vspace{0.05in}
\noindent
Thus, we need a mechanism to eliminate the \io bottlenecks in
transaction processing. \sysname achieves this goal with a new architecture
and a novel authenticated data structure \dsm. With faster transaction
processing, \sysname enables larger blocks \emph{without} changing the block
creation rate. Thus, \sysname increases the throughput of public
blockchains without compromising their security or liveness and without
diluting their decentralization.

%% file: design2.tex
\section{RainBlock}
\label{sec-rainblock}

\input{fig-rainblock}

\sysname aims to achieve the following goals simultaneously:
\begin{itemize}
\item High transaction throughput
\item Scalability with increasing system state and number of users
\item Support for Turing-complete smart contracts~\cite{szabo1994smart}   
\item The same degree of decentralization and security as Ethereum
\end{itemize}  

Achieving these goals simultaneously in public blockchains is challenging.
For example, achieving higher throughput by assuming large amounts
of DRAM at every participating server reduces decentralization as
only specific servers can participate.

\sysname proposes a new architecture for public blockchains that
achieves all these goals. \sysname delivers high throughput by
tackling the \io bottleneck on two fronts. \sysname avoids \io
in the critical path using \emph{prefetching clients} and reduces
\io amplification using the novel \dsm.
We now illustrate the
\sysname architecture and how it achieves scalability and
supports smart contracts, without compromising on
decentralization or security.

\subsection{High-Level Design}
In this section, we buildup the design of \sysname. We
start with the problems that our study on Ethereum highlights. We discuss
how \sysname solves these problems and the resulting challenges.

\subsubsection{Problem-I: \io amplification from storing Merkle trees
  in key-value stores}

Ethereum stores system state in a Merkle tree~\cite{ethereum-mpt},
and persists it using the RocksDB~\cite{rocksdb} key-value store.
Traversing such a Merkle tree requires looking up nodes using their
hashes. Hashing is computationally expensive and results in the
nodes of the tree being distributed to random locations on storage.
As a result, traversing the Merkle tree to read a leaf value
requires several random read operations. The log-structured merge
tree~\cite{o1996log} that underlies RocksDB results in additional
\io amplification~\cite{raju18mlsm,RajuEtAl17-PebblesDB}.

\vheading{Solution: store state in an optimized in-memory
  representation}. \sysname introduces an in-memory version of the
Merkle tree. Persisting the data is done via a write-ahead log and
checkpoints.  Traversing the Merkle tree is \emph{decoupled} from
hashing; obtaining the next node in the tree is a simple pointer
dereference. \sysname introduces a technique termed \emph{lazy hash
  resolution}: when a leaf node is updated in a Merke tree, all the
nodes from the leaf to the root need to be re-hashed; \sysname defers
the re-hashing until the nodes are actually read. Lazy hash resolution
is effective since hashing requires serializing the node contents~\cite{rlp};
thus, lazily hashing the nodes saves both hashing and serialization
operations. Note that simply running RocksDB in memory would not be
effective: the hashing and serialization would still add
significant overhead. 

\subsubsection{Resulting challenge: Tackling Scalability and Decentralization}
Simply keeping the Merkle tree in memory does not achieve the goals of
\sysname. As the blockchain grows, the amount of state in the Merkle
tree will increase; soon, a single server's DRAM will not be
sufficient. Furthermore, for maintaining decentralization, we cannot
require servers to have significant amount of DRAM.

\vheading{Solution: decouple storage from servers and shard the state}. \sysname solves this problem using separate storage nodes. \sysname shards the Merkle
tree into subtrees such that each subtree fits in the memory of a storage node. As the
amount of data in the blockchain increases, \sysname increases the
number of shards. In this manner, \sysname scales with commodity
servers and storage nodes
without diluting the decentralization. 

\subsubsection{Problem-II: Miners perform \io in the critical path} 

On receiving a new block, miners in Ethereum process its transactions
by traversing and updating the Merkle tree in RocksDB (causing random
\io on storage) and verifying if their Merkle root matches the Merkle
root in the block. Only then can the miner process the next block of
transactions. Thus, transaction processing includes performing slow
\io operations in the critical path; and the transactions are
processed one at a time.

\vheading{Solution: decouple \io and transaction execution}. \sysname
solves this problem by removing the burden of doing \io from the
miners. \sysname introduces \emph{prefetching clients} (clients) that prefetch data and witnesses
from the storage nodes and submit them to the miners. Miners use this information to execute transactions without
performing \io and asynchronously update the storage nodes. Since transaction processing now becomes a pure CPU
operation, it is significantly faster. This architecture also
increases parallelism as multiple clients can be prefetching data
for different transactions at the same time. 

\subsubsection{Resulting challenge: Prefetching \io for smart contracts}

One challenge with clients prefetching data for transactions is that
some transactions invoke smart contracts. Smart contracts are
Turing-complete programs that may execute arbitrary code. Thus,
how does the client know what data to prefetch? 

\vheading{Solution: pre-execute transactions to get their read and
  write sets}. \sysname solves this problem by having the clients
pre-execute the transactions. As part of this execution, the clients
read data and witnesses from the storage nodes. One challenge is that the
pre-execution may have different results than when the miner executes
the transactions (\eg the smart contract may execute different code
based on the block which it appears in). We will describe how clients
handle smart contracts correctly despite stale data from pre-execution (\sref{sec-spec}).

\subsubsection{Resulting challenge: Consistency in the face of concurrency}

Another challenge that arises due to the \sysname architecture is
consistency. Multiple clients are reading from the storage nodes, and
multiple miners are updating them in parallel. Using locks or other similar mechanisms will reduce concurrency and throughput.

\vheading{Solution: store system state in the two-layer,
  multi-versioned \dsm}. The \dsm is an in-memory, sharded,
multi-versioned Merkle tree. \sysname uses \dsm to store the system state.
The \dsm has two layers:
\begin{itemize}[leftmargin=*]
\item The \emph{bottom layer} is sharded across the storage nodes and
  contains multiple versions. Every write causes a new version to be
  created in a copy-on-write manner; there are no in-place updates. As a result, concurrent
  updates from miners simply create new versions and do not conflict with each
  other. When a fork of the blockchain is discarded, the bottom layer
  garbage collects the associated versions.
\item The \emph{top layer} represents a consistent version of
  the tree. Each miner has a top layer that is private to the
  miner. The miner executes all transactions against the data and
  witnesses in its top layer. New versions being created in the bottom
  layer do not affect the version in the top layer, ensuring
  consistency. 
\end{itemize}  

\subsubsection{Resulting challenge: \sysname has higher network traffic}

Finally, the architecture of \sysname trades local disk \io for remote
network \io. As a result, \sysname results in more network
utilization, and the network bandwidth may become the bottleneck.

\vheading{Solution: \sysname reduces network \io via deduplication and
  the synergy between the \dsm layers}. \sysname uses multiple
optimizations to reduce network \io. First, the bottom and top layer
of \dsm collaborate with each other; when the bottom layer sends
witnesses to the top layer, it will skip sending nodes of the Merkle
tree that are known to be present at the top layer. We term this
\emph{witness compaction}. Second, when any component of \sysname
sends witnesses over the network, it will batch witnesses and perform
deduplication to
ensure only a single copy of each Merkle tree node is sent. We term
this \emph{node bagging}. Finally, miners send logical updates to storage
nodes rather than physical updates as logical updates are smaller in size. 

\input{arch}

\input{client}

\input{benefits}

%% file: fig-rainblock.tex
\begin{figure*}[t]
  \centering
  \begin{minipage}[b]{.31\textwidth}\ContinuedFloat
      \captionsetup{labelformat=empty}
      \includegraphics[width=\columnwidth,height=5cm]{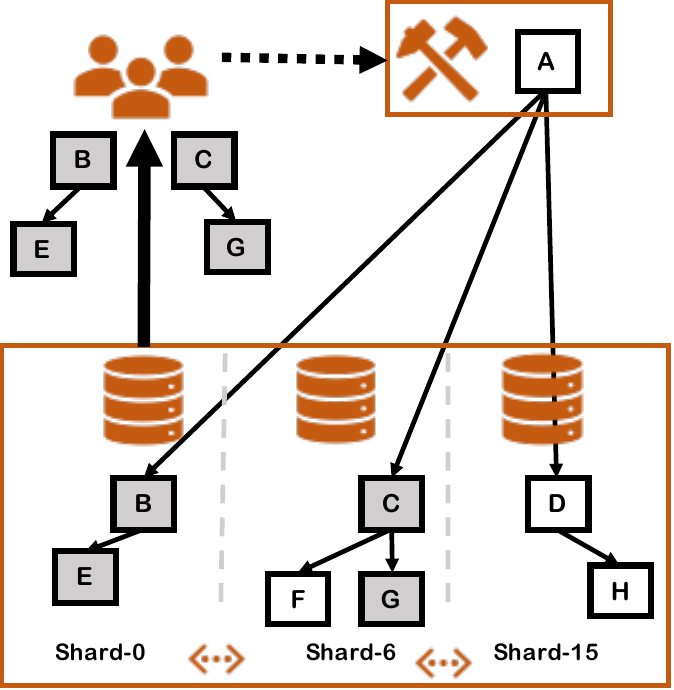}
      \caption{(A) Clients prefetch compact witnesses}
  \end{minipage}
  \hfill{\color{lightgray}\vline}\hfill
  \begin{minipage}[b]{.29\textwidth}\ContinuedFloat
      \captionsetup{labelformat=empty}
      \includegraphics[width=\columnwidth,height=5cm]{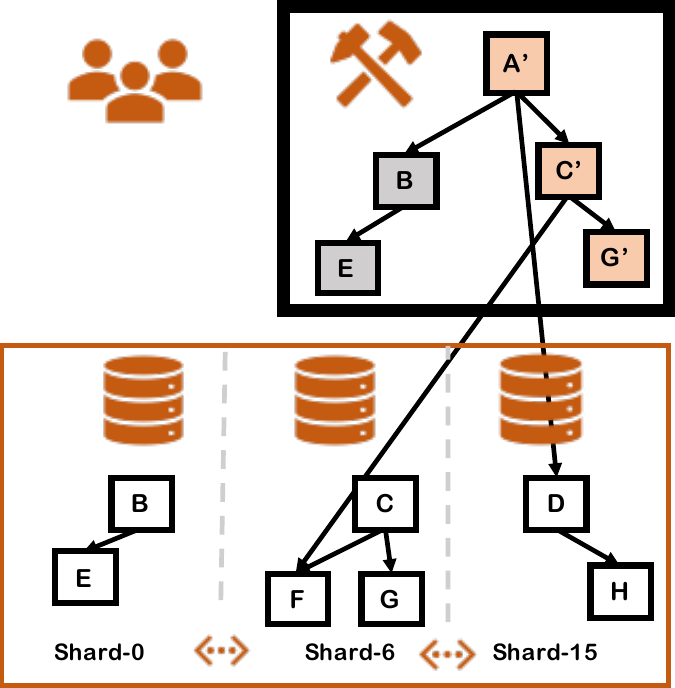}
      \caption{(B) Miners execute txns without \io}
  \end{minipage}
  \hfill{\color{lightgray}\vline}\hfill
  \begin{minipage}[b]{.29\textwidth}\ContinuedFloat
      \captionsetup{labelformat=empty}
      \includegraphics[width=\columnwidth,height=5cm]{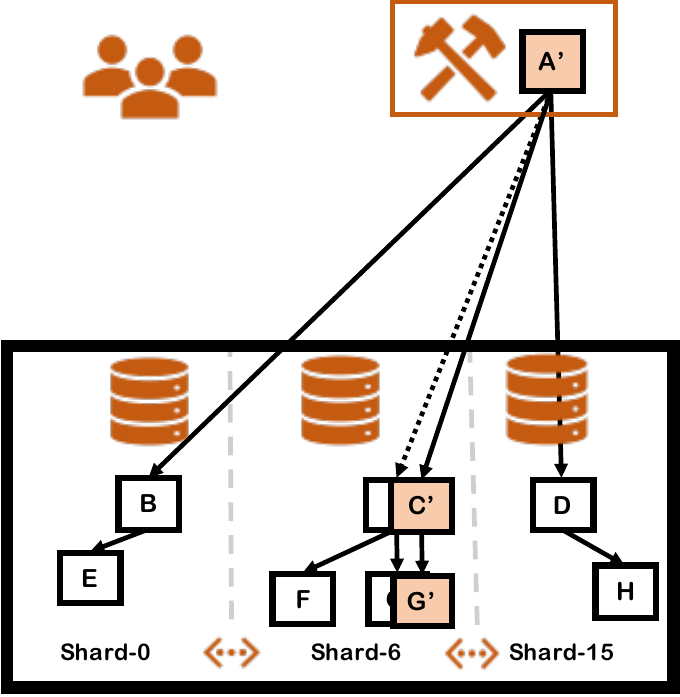}
      \caption{(C) Shards update asynchronously}
  \end{minipage}
  \mycaption{\sysname architecture}{\sysname processing a $Txn$ that reads and updates accounts in two shards that are along the paths $ABE$ and $ACG$. (A) Clients prefetch compact witnesses $BE$ and $CG$ from storage nodes and submit to miners. (B) Miners verify and use these witnesses to execute $Txn$ against their top layer, and later update storage nodes. (C) Storage nodes verify updates from miners and asynchronously update their bottom layer, creating a new version for the modified account $A'C'G'$.}
    
    \label{fig:architecture}
\end{figure*}

%% file: arch.tex
\subsection{Architecture}
\label{sec-arch}

We now describe the \sysname architecture in
detail.

\vheading{Overview}.
\sysname introduces three kinds of participating entities:
\emph{prefetching clients} (clients), \emph{miners}, and
\emph{storage nodes}. Users send transactions to clients.
Clients pre-execute these transactions and prefetch data
and witnesses from storage nodes. Clients submit transactions
and the prefetched information to miners, Figure-\ref{fig:architecture}(a).
Miners are responsible for creating new blocks of transactions
and extending the blockchain; each miner maintains a private copy
of the top layer of the \dsm. Miners use the submitted information
to execute these transactions against their top layer,
Figure-\ref{fig:architecture}(b); and do not perform \io in the
common case. Finally, miners create a new block, gossip it to
other miners, and update the storage nodes. Storage nodes are
responsible for maintaining and serving the system state. They use the
multi-versioned bottom layer of the \dsm to provide consistent
data to clients while handling concurrent updates from miners.
Storage nodes asynchronously update the bottom layer of
the \dsm, Figure-\ref{fig:architecture}(c).



\input{life}

%% file: life.tex
\vheading{The Life of a Transaction in \sysname}.
We now describe the various actions that take place from the time a
transaction is submitted, to when it becomes part of the blockchain.

\begin{enumerate}
\item A user submits the transaction (tx) to a client. 
\item The client will pre-execute the transaction, reading data and
  witnesses from storage nodes
\item The client will optimize the witnesses before sending them over
  the network using node bagging
\item The client will submit the transaction, data, and optimized
  witnesses (node bags) to miner
\item The miner will verify these node bags and advertise them to other miners
\item The miner will execute tx using witnesses + top layer of the
  \dsm. The miner does not need to perform any \io
\item The miner creates the new block, and sends it to other miners
\item The miner sends new block + updates to storage nodes as logical
  operations (\eg A->A') with new Merkle root
\item The storage nodes verify if block is valid (proof of work
  check), log updates, return successful to miner
\item The storage nodes apply the updates and verify them (based on provided Merkle root)
\item The other miners verify the block using their top layer and
  node bags without any \io, and gossip to other miners
\item The tx is added to the blockchain when its block is processed by
  majority of miners
  
\end{enumerate}

%% file: client.tex
\subsection{Speculative Pre-Execution by Clients}
\label{sec-spec}


In \sysname, clients read all the witnesses required for executing a
transaction from storage nodes. These transactions can be simple or can call
smart contracts.
Since smart contracts are Turing-complete, it is not known apriori
what locations they will access. \sysname clients handle this by
\emph{speculatively} pre-executing the smart contract to obtain the
data that is read or modified by the smart contract.

\vheading{Speculative pre-execution}.  Smart contracts can use the
timestamp, or block number of the block in which they appear, during
their execution at the miner. These values are not known yet during
their pre-execution at the clients. As a result, clients speculatively
return a guess while pre-executing the contract. Our analysis of
Ethereum contracts shows that despite providing estimated values,
clients still successfully prefetch the correct witnesses and node
bags.  For example, the CryptoKitties \texttt{mixGenes} function as
shown in Figure~\ref{fig:mixgenes} repeatedly references the current
block number and its hash. Since these numbers are only used to
generate randomness of \emph{written} values in the function,
substituting inaccurate values does not affect the witnesses that are
pre-fetched.

\input{fig-mixgenes}

\vheading{Stale data}. We make a similar observation that clients can
pre-execute with \emph{stale} data and still prefetch the correct node
bags.  For example, the CryptoKitties \vtt{giveBirth} function is a
\emph{fixed-address} contract, where the addresses read (loads from
the \texttt{kitties} array) only depend on the inputs from the message
call. To deal with rare variable-address contracts, the miner may
asynchronously read from a storage node after the transaction is
submitted. Even in these cases, the client will have retrieved some of
the correct witnesses required for the transaction (\eg the \texttt{to}
and \texttt{from} accounts).

\vheading{Benefits from pre-executing clients}. One of the main advantages of
the speculatively pre-executing client is that it can filter out transactions
that miners would abort.  Contrast this with the Ethereum blockchain,
in which aborted transactions are no-ops, but still take up valuable
space in the public ledger.  In \sysname, clients can prevent miners
from spending valuable cycles executing transactions that will be
eventually aborted. The tradeoff is that staleness might cause clients
to abort transactions conservatively.
If users believe that a client incorrectly aborted their transaction, they can send it to other clients or prefetch the witnesses themselves.

%% file: fig-mixgenes.tex
\begin{figure}
\begin{algorithmic}[1]
\hrule
\footnotesize
\vspace{5pt}
\Function{mixGenes}{mGenes, sGenes, curBlock}
\State uint256 $\textit{randomN} \gets curBlock.blockHash$
\State $\textit{randomN}\gets KeccakHash(randomN, curBlock)$
\State MemoryAry $\textit{babyGenes}\gets mix(mGenes, sGenes, randomN)$
\State \textbf{return} $babyGenes$
\EndFunction
\vspace{5pt}
\hrule
\end{algorithmic}
\vspace{-5pt}
    \mycaption{Indeterminate contract values do not affect
        pre-fetching}{Psuedocode of CryptoKitties
      $mixGenes$ function. It makes repeated calls to
      $curBlock$. Although client substitutes it with a
      speculative value, it doesn't
      affect witness prefetching because these numbers only affect
      $written$ values.}
    \label{fig:mixgenes}
\end{figure}

%% file: benefits.tex
\subsection{Benefits}

In summary, \sysname achieves high throughput by reducing \io amplification (using \dsm to scalably store the system state) and eliminating \io bottlenecks (using clients to decouple \io from transaction execution). Thus, \sysname increases the transaction processing rate at miners without assuming any hardware limits on them. With faster transaction processing, \sysname allows miners to pack more transactions per block without impacting the underlying PoW consensus (\sysname does not impact block creation rate). Thus, \sysname achieves high throughput without compromising the security or decentralization of public
blockchains.

The \sysname architecture has many additional benefits:
\begin{itemize}[leftmargin=*]
\item Every component is typically visited once for processing
  a transaction, presenting an efficient architecture for public
  blockchains.
\item \sysname can scale with increasing transaction load (by
  increasing number of clients) and with increasing system state (by
  increasing number of storage shards).
\item Clients can execute read-only transactions bypassing miners.
\item \sysname supports transactions on the sharded system state
  without requiring locking or additional coordination among clients
  and miners.
\item \sysname does not assume trust between any of the components.  
\end{itemize}

%% file: dsm3.tex
\section{\Dsmtree}
\label{sec-dsm}

The \dsmfull (\dsm) is an in-memory, multi-versioned, sharded variant
of the Merkle tree. The \dsm has two layers; we first present the
common in-memory representation, then describe each layer in turn, and
then discuss how the layers collaborate and their trade-offs in different
configurations.

\subsection{In-Memory Representation}

\dsm uses an efficient in-memory representation of the Merkle
tree. Tree traversal is decoupled from hashing: traversing the Merkle
tree is done by dereferencing pointers; in contrast, Ethereum's Merkle
tree has to perform expensive cryptographic hashing to find the next
node during traversal. \dsm uses periodic
checkpoints for persisting the data. The checkpoints are only
used to reconstruct the in-memory data structure in case of failures;
reads are always served from memory.

\vheading{Lazy Hash Resolution}. When a leaf node in a Merkle tree is
updated, hashes of nodes from the leaf to the root need to be
recomputed. \dsm defers doing this recomputation until a node is
actually read. This makes writes efficient as only the leaf node has
to be updated in the critical path. Recomputing hashes is expensive as
nodes have to be serialized before being hashed; as a result, lazy
hash resolution improves performance significantly by saving expensive
hashing and serialization operations~\cite{rlp}. 

\input{fig-dsm}

\subsection{Bottom Layer}

The bottom layer of the \dsm consists of a number of shards. Each
shard is a vertical subtree of the merkle tree, stored in DRAM. The
bottom layer supports multiple versions to allow concurrent updates to
the \dsm, as shown in Figure-\ref{fig:dsm-layers}. The bottom layer has a
write-ahead log to persist logical updates.

\vheading{Multi-versioning}. Each write to the bottom layer creates a
new version of the tree. There are no in-place updates. This versioning is
required as miners may submit multiple blocks concurrently that
potentially conflict with each other; the bottom layer creates a new
version for each write. It creates versions only for the modified data in a copy-on-write manner. Thus, writes never conflict with each other,
and \dsm does not require locking or additional coordination among
miners or clients.

\vheading{Garbage collection}. Garbage collection of versions is
driven by the higher-level blockchain semantics. When multiple miners
are working on competing forks of the blockchain, multiple versions
are maintained. Eventually, one of the forks is accepted as the
mainline fork, and the others are discarded (and their associated
versions are garbage collected by the bottom layer of the \dsm).

\subsection{Top Layer}

Given that the bottom layer maintains multiple versions across
multiple shards, we need a way for miners to access data in a
consistent fashion. The top layer provides this mechanism.

Each miner has a private top layer. The top layer contains the first
few levels of the Merkle tree, till a configurable retention level
($r$). The top layer has the Merkle root node that summarizes the entire
system state, and presents a consistent snapshot of the system
state. Miner executes transactions against this snapshot; all
reads return values from this snapshot of the system.

As the miner executes transactions, their top layer is updated,
switching to a different consistent view of the system state,
as shown in Figure-\ref{fig:dsm-layers}. The
changes in the top layer's Merkle tree will also be reflected in the
bottom layer's storage shards after the miner sends logical updates
to the storage nodes.

\vheading{Caching and Pruning}.  The top layer acts as an in-memory
cache of witnesses for the miners.
By design, the top layer stores the
recently used and the frequently changing parts of the Merkle tree.
The top layer receives witnesses
(or paths of the Merkle tree) from the prefetching clients. The top
layer uses the node bags from the clients to reconstruct a
\emph{partial} Merkle tree that allows miners to execute transactions,
typically without performing \io from the bottom layer. The top layer
also supports pruning the partial Merkle tree to help miners reclaim
memory. Pruning replaces the nodes at the retention level ($r$ + 1)
with \emph{Hash nodes}. Hash nodes also help miners to identify the
\dsm shard which has pruned nodes.

\vheading{Witness Revision}. In \sysname, the bottom layers of the
\dsm update asynchronously. Therefore, the top layer (miner) may
receive stale witnesses from the bottom layer (prefetching clients).
\dsm introduces a new technique termed \emph{witness revision} to
tolerate stale witnesses. A witness is determined to be stale or
incorrect because the Merkle root in the witness doesn't match the
top layer's Merkle root. However, this could happen because of an
\emph{unrelated update} to another part of the Merkle tree. The top
layer detects when this happens, and \emph{revises} the witness to
make it current. If the Merkle root matches now, then the witness
is accepted. Witness revision is similar to doing \vtt{git push}
(trying to upload your changes), finding out something else in
the repository has changed, doing a \vtt{git pull} (obtaining
the changes in the repository) to merge changes, and then doing
a \vtt{git push}. With witness revision, the top layer tolerates
stale data from the bottom layer and allows miners to execute
non-conflicting transactions that would otherwise get rejected.
Note that, witness revision cannot revise every potential stale
witnesses. If the top layers are pruned aggressively, the top
layer may have insufficient information to detect if the changes
are from an unrelated part of the Merkle tree.

\subsection{Synergy among the layers}

The top and bottom layers collaborate to reduce network traffic. We also briefly
discuss the potential \dsm configurations with $r$ (retention at the top layer)
and $c$ (compaction level at the bottom layer), and the tradeoffs involved.

\vheading{Witness compaction}.  As the top layer of the \dsm stores
the top levels of the Merkle tree, the storage nodes do not need to
send a full witness. Like the configurable retention level at the top
layer $r$, the bottom layer has a configurable compaction level,
$c$. Only nodes below the compaction level (compact witnesses)
are sent in node bags, after deduplicating nodes across witnesses,
reducing the network burden of transmitting witnesses.

\vheading{Configurations}.
The \dsm is configurable to operate entirely from local memory without any network overhead, or just from remote memory with high network utilization. For example, If the top layer of the \dsm has $r=$ $\infty$, then the top layer caches the entire Merkle tree and is fully served from local memory. Similarly, if the bottom layer has $c=0$, then un-compacted witnesses are sent over the network and accessed entirely from remote memory. Thus, \dsm provides a unique (and flexible) point in the design spectrum of distributed, in-memory, authenticated data structures.

\vheading{Tradeoffs}.  
In a Merkle tree that has $n$ levels, any \dsm configuration that satisfies $c >= (n-r)$
allows the top layer to use compacted witnesses from the bottom layer. Note that having a higher
$r$ results in a lower number of transaction aborts, as the top layer has more
information to detect non-conflicting updates and perform witness
revision. Therefore in \sysname, ideally, the top layers should set $r$ based on
the amount of memory available. Pruning the top layer should only be done under
memory pressure.

\subsection{Summary}

In summary, the \dsm is a novel variant of the Merkle tree modified
for faster transaction processing in public blockchains. It uses an
efficient in-memory representation to reduce \io amplification,
multi-versioning to handle concurrent updates, and the Merkle root
in the top layer to
provide a consistent view of the system state. \dsm presents a new
point in the design space of authenticated data structures. The top
layer exploits the cache-friendliness of the Merkle tree, while the
sharded bottom layer relies on the fact that witness creation only
requires a vertical slice of the tree. While the \dsm is exclusively
used with \sysname in this paper, it can be easily modified to work
with other blockchains.

%% file: fig-dsm.tex
\begin{figure}[t]
  \centering
  \begin{minipage}[b]{0.7\columnwidth}\ContinuedFloat
      \captionsetup{labelformat=empty}
      \caption{\dsm top layer at miners}
      \includegraphics[width=\columnwidth]{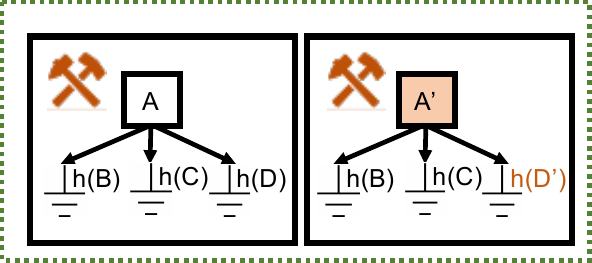}
  \end{minipage}\\
  \begin{minipage}[b]{0.9\columnwidth}\ContinuedFloat
      \captionsetup{labelformat=empty}
      \includegraphics[width=\columnwidth]{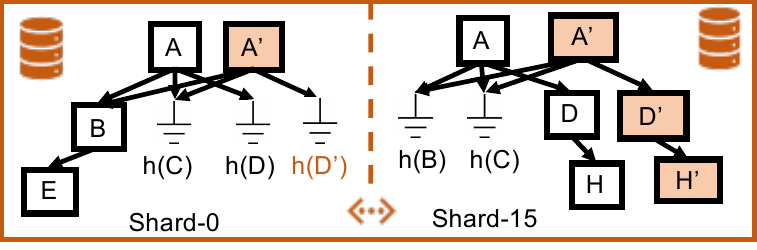}
      \vspace{-20pt}
      \caption{\dsm bottom layer sharded across storage nodes}
  \end{minipage}
  \mycaption{
      \dsm design in \sysname}{This figure shows the two-layered \dsm where miners have their private copy of the top layer for consistency and the bottom layer is sharded for scalability.}
  \label{fig:dsm-layers}
\end{figure}

%% file: discussion.tex
\section{Discussion}
\label{sec-discussion}

We now discuss the trust assumptions, incentives, security, and
limitations of the \sysname architecture. 

\vheading{Trust Assumptions}. In keeping with existing public
blockchains, \sysname \textbf{does not require trust} between any of
its components. Miners operate without trusting the clients or the
storage nodes by re-executing transactions and verifying the data they
receive. Clients operate without trusting the storage nodes and
miners, as clients verify their reads from storage nodes, and can
verify the block produced by a miner. Finally, storage nodes also
operate without trusting miners, as they can verify updates from
miners.

\vheading{Incentives}. We sketch a possible incentive model here. A
more rigorous analysis would require game-theoretic and economic
models, which are beyond the scope of this paper.  Users can prefetch
data from the storage nodes themselves or pay clients. Users, clients,
and miners pay the storage nodes for reading authenticated
data. Miners get paid for mining a new block of transactions through
block rewards. Every \sysname component can detect misbehaving
entities and blacklist them, incentivizing correct behavior.

\vheading{Security}. \sysname provides the same security
guarantees as Ethereum, as it does not change the consensus protocol
(Proof of Work) or trust assumptions between participating servers.
Further, \sysname does not impact the block creation rate by
packing more transactions into each block.


\vheading{Limitations}. \sysname trades local storage \io for network
\io, so the network may become a bottleneck. \sysname recognizes this
risk and uses multiple techniques such as witness compaction and node
bagging to reduce network traffic.


%% file: impl.tex
\section{Implementation}
\label{sec-impl}

We implement \sysname and \dsm in Typescript, targeting node.js. The
miners and storage nodes use the \dsm as a library. The performance
critical portions of the code, such as \texttt{secp256kp1} key
functions for signing transactions and generating \texttt{keccak}
hashes, are written as C++ node.js bindings. To execute smart
contracts, we implement bindings for the Ethereum Virtual Machine
Connector interface (EVMC) and use Hera (v0.2.2). Hera can run
contracts implemented using Ethereum flavored WebAssembly (ewasm) or
EVMC1 bytecode through transcompilation. Our speculative pre-executing
client is implemented in C++.  The \dsm and \sysname implementations,
together 15K lines of code, is open source and available on
GitHub\footnote{https://github.com/RainBlock}. Our current
implementation of storage nodes assumes a 16-way sharded Merkle
Patricia tree by default. It supports a configurable number of
shards.

%% file: eval.tex
\section{Evaluation}
\label{sec:eval}

In this section, we evaluate the performance of \dsms and \sysname.
We seek to answer the following questions:
\begin{enumerate}
\item What is the performance of the \dsm for various operations on a
  single node? (Section-\ref{sec:micro})
    \item How does the size of the \dsm cache impact the witness
      sizes, memory overhead, and rate of transaction aborts in \sysname?
      (Section-\ref{sec:tuning})
  \item What is the performance of \sysname on various end-to-end
    workloads that characterize the Ethereum public blockchain?
    (Section-\ref{sec:e2e})
\end{enumerate}

\subsection{Experimental Setup}
\label{exp-setup}

We run the experiments in a cloud environment on instances which are
similar to the \texttt{m4.2xlarge} instance available on Amazon EC2
with 32GB of RAM and 48 threads per node. We use Ubuntu 18.04.02 LTS,
and node.js v11.14.0.  For the end-to-end benchmarks, each storage
node, miner, and client is deployed on its own instance.

\subsection{Evaluating \dsm on a single node}
\label{sec:micro}

\input{fig-comb-micro.tex}

First, we evaluate the \dsmtree running on a single node. This tests
the performance of the optimized in-memory representation of \dsm. We
measure the throughput of point \texttt{put} and \texttt{get}
operations for a variety of tree sizes against the state-of-the-art
Ethereum MPT.
Point \texttt{put} operations create or update a key-value pair and \texttt{get} operation returns the value and witness for a key.

To make a fair comparison, we compare \dsm with the in-memory
implementation of Ethereum MPT~\cite{ethMpt}. The in-memory Ethereum MPT uses
\texttt{memdown}~\cite{memdown}, an in-memory key-value store built on a
red-black tree. We are comparing an in-memory MPT that uses the key-value representation to the in-memory
\dsm. The difference in performance comes from the in-memory
design and optimizations in \dsms, and not due to different
storage media.

We dump the Ethereum world state every 100K
blocks until 4M blocks and use it to micro-benchmark \dsms;
every key in these benchmarks is a 160-bit Ethereum address and values
are RLP-encoded Ethereum accounts~\cite{rlp}.

\vheading{Gets}. \dsms with 1.19M accounts, obtain a \texttt{get}
throughput of $\approx$\textbf{216K ops/s}, that is \textbf{150\myx}
the throughput of Ethereum MPT. The main reason for the \dsm's better
performance is the use of in-memory pointers. To fetch a node, the
\dsm simply needs to follow a path of in-memory pointers to the leaf
node.  On the other hand, walking down a tree path means looking up
the value (node) at a particular hash for each node in the Ethereum
MPT.  Even though this database is in-memory, looking up values in an
in-memory key-value map is still more expensive than a few pointer
lookups.  Furthermore, the larger the world state, the better \dsm's
in-memory Merkle tree performs over the Ethereum MPT. This is simply
because the larger the state, the taller the tree, so the more nodes
on the path to a leaf, see Figure~\ref{fig:micro} (a).

\vheading{Puts}. \dsms with 1.19M accounts obtain a \texttt{put}
throughput of $\approx$\textbf{245K ops/s}, that is \textbf{160x} the
throughput of Ethereum MPT.  Due to lazy hash resolution, a
\texttt{put} does not need to adjust any values in the path from the
leaf to the root; in contrast, every node in the path has to be
updated in the Ethereum MPT. \texttt{put} throughput in the \dsm is
more than two orders of magnitude higher than in the Ethereum MPT.

\vheading{Tree Size}. Figure~\ref{fig:micro} (b) shows that \dsms are
significantly smaller than Ethereum MPTs when the same number of
accounts are stored. With 1.19M accounts, the Ethereum MPT consumes
$\approx$26021MB and \dsms consume $\approx$775MB, using
\textbf{34\myx} lesser memory. The primary reason for this is the
efficient in-memory representation of \dsms. Ethereum MPT is
not-memory efficient as it uses 32-byte hashes as pointers and relies
on \texttt{memdown}~\cite{memdown} to store the flatenned MPT as
key-value pairs.  The significantly reduced size of the \dsm, along
with sharding, enables \dsms to be stored entirely in memory,
eliminating the IO bottleneck.

\vheading{Lazy hash resolution}.  We run an experiment where we
trigger a root hash calculation after every $N$ write (\texttt{put} or
\texttt{delete}) operations.  As $N$ increases, the performance of
\dsm write operations also increases. At $N=1000$ (the root hash is
read every 1000 writes), \dsm is 4--5\myx faster than Ethereum
MPT. Since the root hash calculation is expensive (requiring RLP
serialization of nodes), performing it even once every 1000 writes
reduces \dsm performance from 150\myx Ethereum MPT performance to
5\myx.

\subsection{Impact of cache size}
\label{sec:tuning}

\input{fig-tuning}

Next, we evaluate the performance of the distributed version of the
\dsm when the cache size (retention level of the top layer) is changed. Pruning the cache reduces memory
consumption but results in larger witnesses being transmitted, and
more transactions being aborted due to insufficient witness
caching. We evaluate these effects.

\vheading{Memory consumption}.  We evaluate the reduction in the
application memory utilized, from pruning the \dsm cache, across
varying cache sizes $r$.  Figure~\ref{fig:tuning} (a) shows that lower
$r$ will result in higher memory savings, with a tree of depth five
consuming only 40\% of the memory consumed by the full tree.  However,
this means that either 1) \dsm shards will have to provide larger
witnesses or 2) the application will experience a higher abort rate
due to insufficient witness caching.

\vheading{Witness Compaction}. \dsms transmit compact witnesses which
include only the un-cached parts of the witness. \dsms employ
node-bagging where they combine multiple witnesses and eliminate
duplicate nodes. Figure~\ref{fig:tuning} (b) shows the reduction in
witness size due to node bagging and witness compaction, based on the
height of the cached tree $r$. Witness compaction and node bagging
together reduce witness sizes by up-to \textbf{95\%} of their original
size.

\vheading{Transactions}. Pruning the cache discards cached
witnesses. Since transactions abort if the witnesses are not cached,
this increases the abort rate. To study the effect of varying the
cache size on transaction abort rate, we use \sysname with 16 storage
nodes, 1 miner, and enough clients to saturate the miner.
Transactions are generated by selecting two random accounts from a set
of $N$ accounts.  Figure~\ref{fig:tuning} (c) shows that the
transaction abort rate is dependent on two factors: the \dsm cache
retention level, and the number of accounts.  In particular,
increasing $r$ reduces the transaction abort rate.  More importantly,
with large number of accounts $N$, the contention on Merkle tree nodes
reduces, reducing the abort rate for fixed a $r$, making \dsms
practical for application with low available memory.

\subsection{End-to-End Blockchain Workloads}
\label{sec:e2e}

\input{fig-e2e}

Finally, we evaluate the end-to-end performance of \sysname against
synthetically generated workloads that mirror transactions on the
Ethereum public \emph{mainnet} blockchain.

\vheading{Challenges}.  Since Ethereum transactions are signed, the
public transactions are not conducive to experiments: we cannot change
transaction data or the source accounts, because we do not have the
\texttt{secp256k1} private key. Since \sysname runs transactions at a
much higher rate than Ethereum, we quickly run into state mismatch
errors, and eventually, exhaust the available transactions.

To tackle this challenge, we analyze the public blockchain to extract
salient features, and develop a \emph{synthetic workload generator}
which generates accounts with private keys we control so our clients
can run and submit signed transactions.

\vheading{Synthetic Workload Generator}. We analyze the transactions
in the Ethereum \vtt{mainnet} blockchain to build a synthetic workload
generator.  We analyzed 100K recent (since block 7M) and 100K older
blocks (between blocks 4M and 5M) in the Ethereum blockchain to
determine: 1) the distribution of accounts involved in transactions,
2) what fraction of all transactions are smart contract calls.  We
observe that 10-15\% of Ethereum transactions are contract calls and
the rest are simple transactions. This is true of both recent blocks
and older blocks.  It is also the case that a small percentage of
accounts are involved in most of the transactions.
Based the analyzed data, we generate workloads where 90\% of accounts
are called 10\% of the time, and 10\% of the accounts are called 90\%
of the time. Smart contracts are invoked 15\% of the time.

\vheading{Throughput}.  Figure~\ref{fig:e2e} (a) shows the transaction
throughput results.  First, this figure shows that the \sysname can
achieve an end-to-end verification throughput of 30,000 transactions
per sec.  It also demonstrates the scalability of the \dsm and
\sysname, which scales as more clients are added.  By varying the \dsm
retention level at the miners from 0 to 6, the \dsm shard throughput
increases by \textbf{7x}, from 1.3K ops/s to 9.4K ops/s, increasing
the scalable creation and transmission of witnesses.

\vheading{Geo-distributed Experiment}. We also ran a geo-distributed
experiment, with varying numbers of regions across 3 continents. Each
region has 4 clients, 1 miner, and 16 storage nodes, caching eight
levels of the \dsm tree ($r=8$).  Figure~\ref{fig:e2e} (b) reports the
throughput experienced by the \sysname. \sysname in a single region
achieves a throughput of $\approx$25K transactions/sec; when we scale
to four regions, the throughput drops to $\approx$20K
transactions/sec, thus retaining 80\% of the performance in a
geo-distributed setting.

\vheading{Contract Calls}. We also ran a workload where accounts
repeatedly call the \vtt{OmiseGO Token}, which is an ERC-20 token
contract~\cite{erc20}. Four clients repeatedly called the token
contract against a single \sysname miner with \dsm cache configured at
$r=8$, achieving a throughput of $17.9K\pm 796$ contract calls per
second.  This demonstrates that even for pure contract contract calls,
\sysname can provide orders of magnitude higher transaction throughput
than other blockchains.

%% file: fig-comb-micro.tex
\begin{figure}[!t]
  \includegraphics[width=\columnwidth]{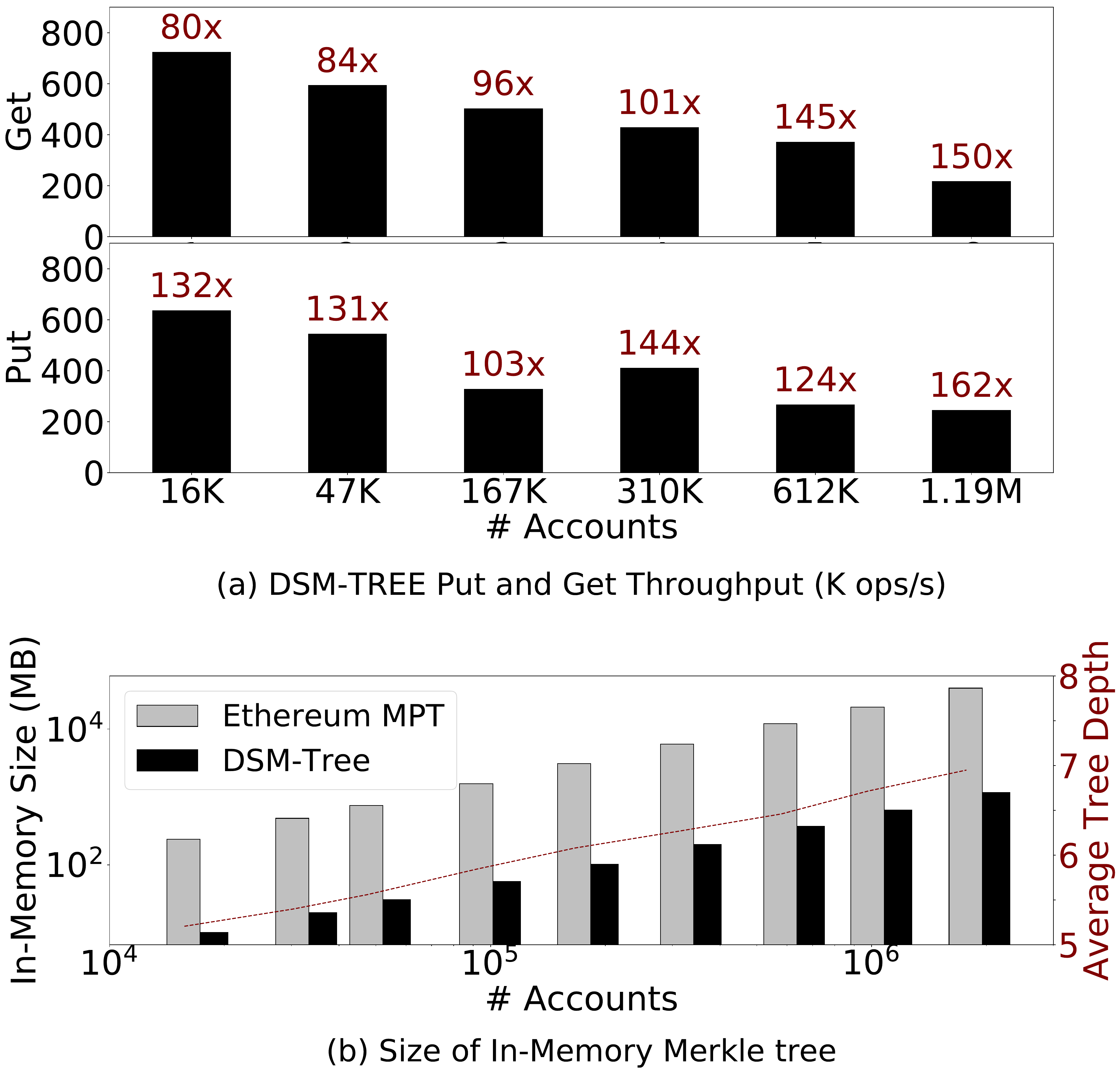}
  \mycaption{Performance of DSM-Tree on a single node}{ (a) The figure
    shows the absolute put and get throughput of \dsms.  Throughput
    relative to the Ethereum MPT is shown on the bars. As the number
    of accounts increase, \dsm throughput increases relative to
    Ethereum MPT.  (b) This figure shows the memory used by \dsm and
    Ethereum MPT across varying number of accounts. The trend line
    captures the height of the MPT.  \dsms are orders of magnitude
    more memory-efficient than Ethereum MPT.  Note the log scale on
    the axes.}
    \label{fig:micro}
\end{figure}

%% file: fig-tuning.tex
\begin{figure}[h!]
    \includegraphics[width=\columnwidth]{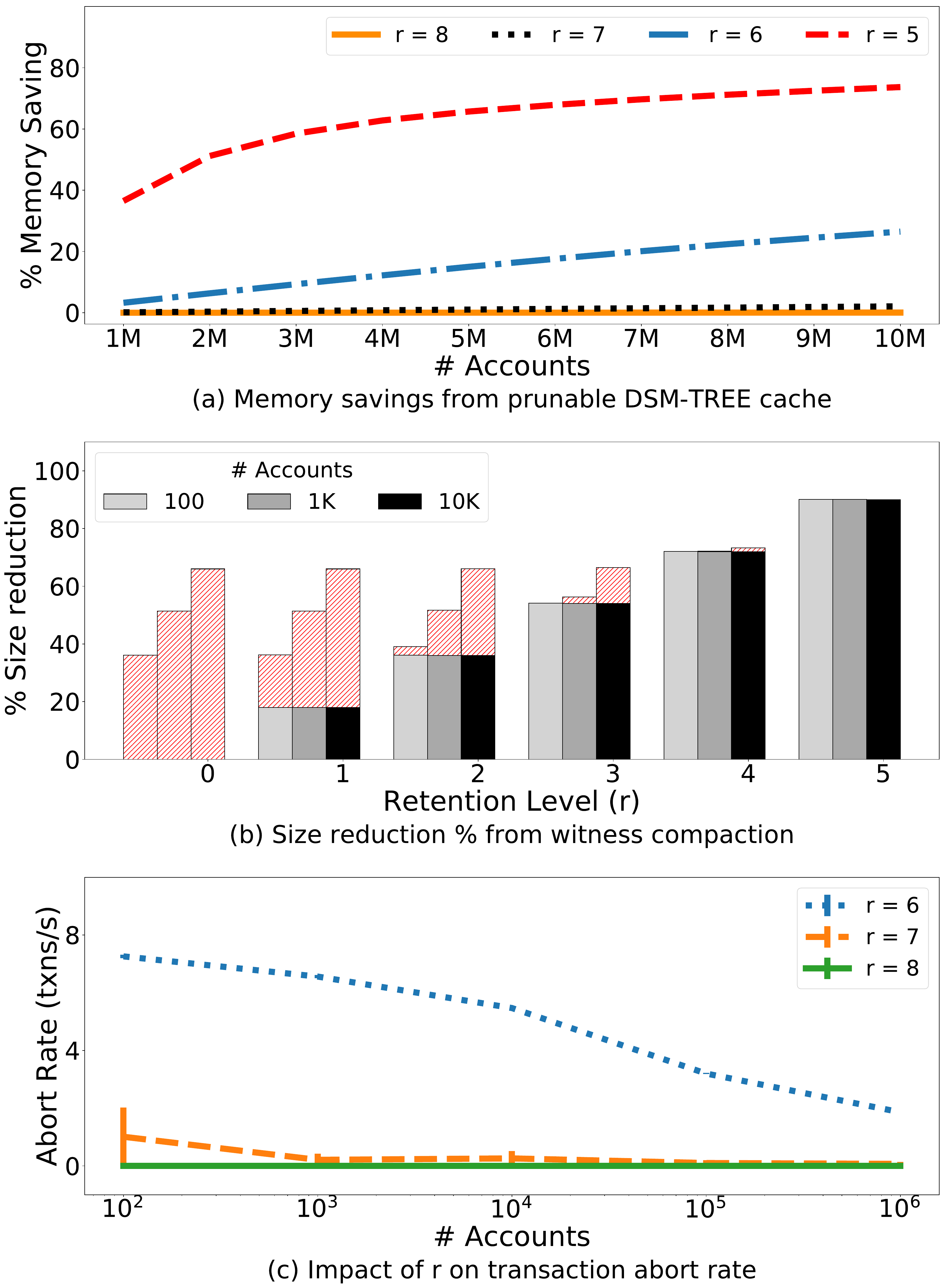}
    \caption{\textbf{Tuning \dsm Cache Retention $r$}. \emph{ (a) This
        figure shows that the caching fewer levels in the cache leads
        to higher memory savings (compared to storing the full tree in
        memory). We do not report the memory savings of higher values
        of $r$ as they were negligible.
        (b) The figure shows the reduction in witness size due to
        combining witnesses and eliminating duplicates (red striped
        bar) and due to witness compaction (solid bar).  (c) The
        figure shows the impact of $r$ (height of cached tree) on transaction
        abort rate.  Higher $r$ results in lower number of transaction
        aborts.  Abort rate decreases with fixed $r$ as the total
        number of accounts $N$ increases, because this reduces the
        probability that transaction will involve accounts that
        conflict at the pruned levels.}  }
    \label{fig:tuning}
\end{figure}

%% file: fig-e2e.tex
\begin{figure}
  \includegraphics[width=\columnwidth]{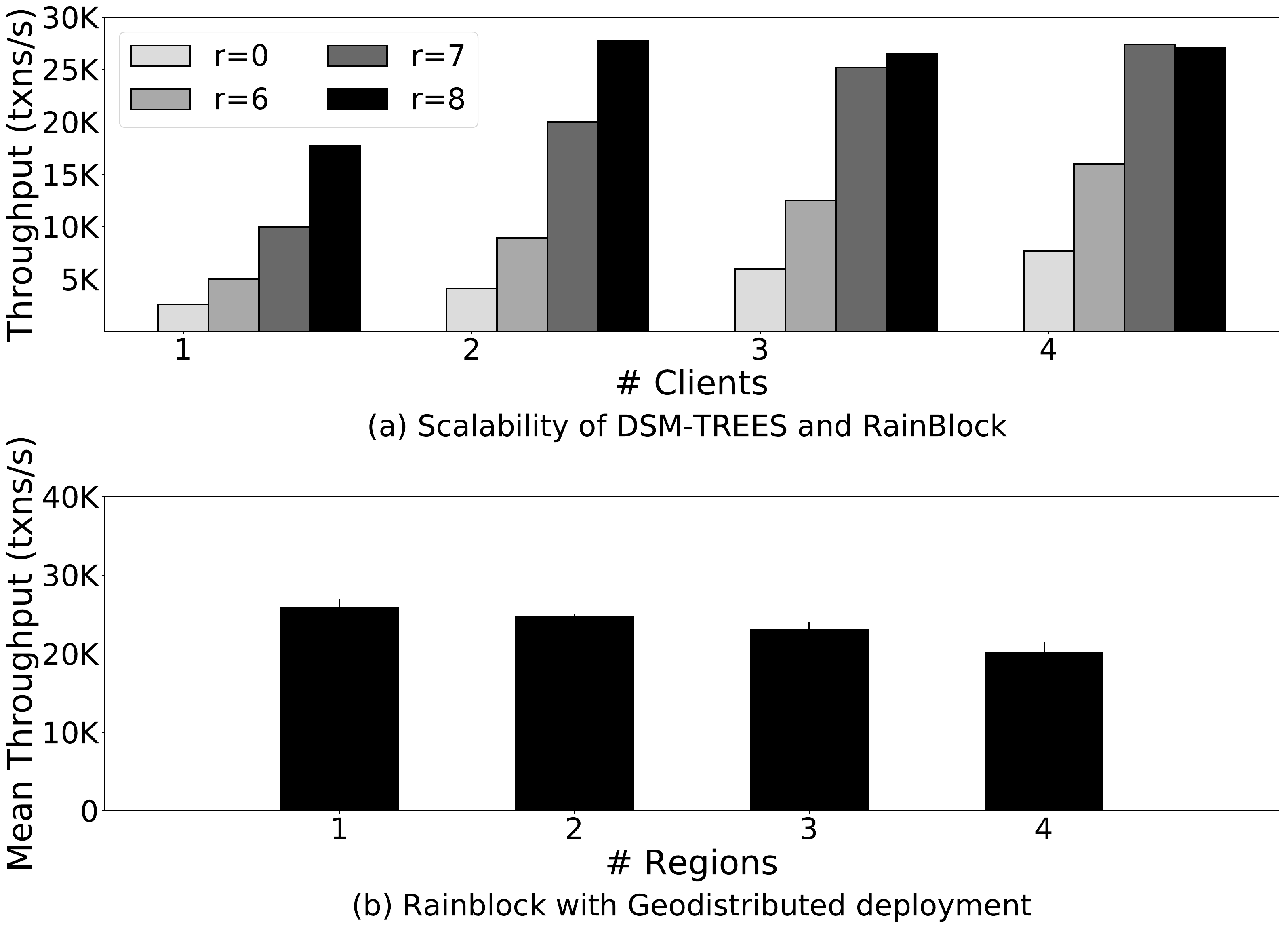}
  \mycaption{End-to-End Blockchain Workloads}{ (a) The figure shows
    the scalability of \dsms in \sysname with increasing number of
    clients and varying cache retention levels ($r$).  The workload
    used in the experiment is representative of the account
    distributions in Ethereum transactions. Miners in \sysname can
    process about about \textbf{30K tps} with 4 clients each, when
    configured at $r = 7$.  (b) This figure shows the overall
    throughput of \sysname in a geodistributed deployment. Miners at
    $r = 8$ can process about \textbf{20K tps} using 4 clients each,
    when communicating with the \dsmtree across WAN.  }
    \label{fig:e2e}
\end{figure}

%% file: related.tex
\section{Related Work}
\label{sec:related}

In this section, we place our contributions \sysname and \dsm, in the context of prior research.

\subsection{Blockchain Systems}

\vheading{Stateless Clients}. The Stateless
Clients~\cite{stateless-clients} proposal seeks to insert witnesses
into blocks, enabling Ethereum miners to process a block without
performing I/O. Despite active
discussions~\cite{ethereum-stateless-solution,
  ethereum-stateless-analysis, ethereum-eip}, Stateless Clients have
not been implemented due to concerns about witness
sizes~\cite{ethereum-stateless-dead-end}. Witnesses for a single,
simple Ethereum transaction can be $\approx$ 4-6KB, resulting in
40-60$\times$ the network overhead. In contrast, \dsm reduces witness
sizes (by $\approx$ 95\%), and \sysname uses prefetching clients to
remove the I/O burden on miners.

\vheading{Private blockchains}. Hyperledger
Fabric~\cite{androulaki2018hyperledger} proposes a novel
execute-order-validate architecture for permissioned (private)
block-chains. Fabric optimistically executes transactions and relies
on the signatures from trusted nodes for verifying transactions. In
contrast, \sysname improves transaction throughput in public
blockchains without trusting any of the participating servers.

\vheading{Sharding and off-chain computation}. Recent work increase
block-chain throughput by sharding the blockchain into independent
parallel chains that operate on subsets of
state~\cite{Plasma,vukolic2015quest,omniledger,Zamani:rapidchain,Luu:Elastico,227661}.
However, sharding requires syncing these independent chains for
consistency, requires complex protocols for cross-shard transactions,
and is less resilient to failures or
attacks~\cite{sonnino2019replay,rajab2020feasibility,yun2019trust}.
In contrast, \sysname does not shard the global blockchain; the
storage is sharded, but all miners add to a single chain. \sysname
does not require locking or additional communication for executing
transactions across multiple storage shards. Payment
channels~\cite{lokhava2019fast,khalil2017revive,green2017bolt,miller2017sprites,Raiden-PaymentChannel,Lightning-Network}
that offload work to side chains are complementary to our work.

\vheading{Consensus}: Ethereum and Bitcoin employ Nakamoto consensus
based on Proof-of-Work (PoW)~\cite{Jakobsson1999ProofsOW,
  sompolinsky2015secure}. There is active research on designing novel
consensus for
blockchains~\cite{,bitcoin-ng,Miller:2016:honeybadger,yin2018hotstuff,cachin2017blockchain,bano2017consensus,203704}
including
Proof-of-Stake~\cite{Gilad:algorand,10.1007/978-3-319-63688-7_12,ethereum-casper} and
Proof-of-Elapsed-Time~\cite{hyperledger-sawtooth} protocols, primarily
because PoW limits block creation rates, trades off wasted work for
security~\cite{state-ddos-attack, 12s-eth}, and is intolerant to the
51\% attack~\cite{eyal2014majority}. While new protocols can replace
PoW in Ethereum, low transaction processing rates will remain a
concern~\cite{yang2019empirically,dinh2017blockbench,10.1145/3319535.3363213,yang2019prism}; since alternative consensus protocols aim at releasing blocks at a higher rate; but perform \io in the critical path for creating new blocks. Therefore, orthogonal to the consensus
protocols, \sysname alleviates \io bottlenecks in transaction processing to increase the throughput of public blockchains.


\vheading{Discussion}: Recent work on reducing network overheads in
block-chains~\cite{10.1145/3341302.3342082,CompactBlocks,ding2019txilm},
and including forks into the main
chain~\cite{li2018scaling,lewenberg2015inclusive}, are orthogonal to
our work. These techniques can be applied to \sysname to further
increase its overall throughput. Other work using trusted hardware for
reducing storage
overheads~\cite{lind2018teechain,10.1145/3299869.3319889} or consensus
protocols~\cite{behl2017hybrids,10.1145/3007788.3007790} is orthogonal
to our work.

\subsection{Authenticated Data Structures}

\vheading{Dynamic accumulators}. Merkle trees belong to a general
family of cryptographic techniques called dynamic
accumulators~\cite{camenisch2002dynamic,benaloh1993one}. Merkle trees,
known for their fast processing, have proofs that grow with the
underlying state. Constant-size dynamic accumulators based on RSA
signatures~\cite{camenisch2002dynamic, benaloh1993one} have fixed size
proofs. However, constant-size accumulators have low processing rates,
and improving their performance is an ongoing
effort~\cite{boneh2019batching}. \dsm provides a practical solution to
achieve high processing rates and small witness sizes while supporting
transactions.

\vheading{Authenticated data structures}. Recent work has proposed a
number of new authenticated data
structures~\cite{kalidhindiangela,yu2019coded,raju2018mlsm,bailleu2019speicher,reyzin2017improving,xu2019vchain,chepurnoy2018edrax,tomescu2020aggregatable}. In
contrast to these work, \dsm scales Ethereum's Merkle Patricia
trie~\cite{ethereum-mpt} without changing its core structure, or how
proofs are generated.



\subsection{Transactional Stores}

\vheading{Transaction execution}. \sysname adopts a design similar to
Solar~\cite{zhu2018solar} and vCorfu~\cite{wei2017vcorfu}, where
transactions are executed based on data from sharded storage.
\sysname modifies the design for decentralized applications and
authenticated data structures. This allows \sysname to execute
transactions on sharded state without requiring locking or additional
coordination among miners.
The \dsm design argues that large random-access data structures can get higher throughput and scalability when served from memory over the network. While RAMCloud~\cite{ramcloud} proposed a similar idea for lower latency, \dsms employ it for higher throughput.





%% file: conclusion.tex
\section{Conclusion}
\label{sec-conc}

\urlstyle{tt}
\renewcommand\UrlFont{\color{blue}\ttfamily}

\sysname increases the throughput of public blockchains while
achieving the same degree of decentralization and providing strong
security guarantees. \sysname achieves this by tackling the \io
bottleneck in transaction processing. \sysname does not modify the
consensus or affect the block creation rate, and hence does not alter
the security provided by public blockchains such as Ethereum. \sysname
increases the transaction processing rate, thereby enabling larger
blocks and higher throughput.

Please refer to our technical report for more details about
\sysname~\cite{ponnapalli2019scalable}. The \sysname prototype is publicly available at
\texttt{\url{https://github.com/RainBlock}}.
